\documentclass{article} 
\usepackage[final]{colm2026_conference}

\usepackage{wrapfig}
\usepackage{microtype}
\usepackage{hyperref}
\usepackage{url}
\usepackage{booktabs}
\usepackage{amssymb}
\usepackage{amsmath}
\usepackage{multirow}
\usepackage{makecell}
\usepackage{mdframed}
\usepackage{graphicx}

\newcommand{\samethanks}[1][\value{footnote}]{\footnotemark[#1]}

\usepackage{lineno}
\usepackage{enumitem}
\usepackage{subcaption}

\definecolor{darkblue}{rgb}{0, 0, 0.5}
\hypersetup{colorlinks=true, citecolor=darkblue, linkcolor=darkblue, urlcolor=darkblue}

\title{Constraint Decay: The Fragility of LLM Agents in Backend Code Generation}

\author{Francesco Dente\thanks{Equal contribution.}\\
EURECOM, France\\
\texttt{francesco.dente@eurecom.fr} \\
\And
Dario Satriani\samethanks\\
University of Basilicata, Italy\\
\texttt{dario.satriani@unibas.it} \\
\And
Paolo Papotti\\
EURECOM, France\\
\texttt{papotti@eurecom.fr}
}

\begin{document}

\ifcolmsubmission
\linenumbers
\fi
\maketitle

\begin{abstract}
Large Language Model (LLM) agents demonstrate strong performance in autonomous code generation under loose specifications. However, production-grade software requires strict adherence to structural constraints, such as architectural patterns, databases, and object-relational mappings. Existing benchmarks often overlook these non-functional requirements, rewarding functionally correct but structurally arbitrary solutions. We present a systematic study evaluating how well agents handle structural constraints in multi-file backend generation. By fixing a unified API contract across 80 greenfield generation tasks and 20 feature-implementation tasks spanning eight web frameworks, we isolate the effect of structural complexity using a dual evaluation with end-to-end behavioral tests and static verifiers.
Our findings reveal a phenomenon of constraint decay: as structural requirements accumulate, agent performance exhibits a substantial decline. Evaluated configurations lose 27.28 points on average in assertion pass rates from baseline to fully specified tasks.
Framework sensitivity analysis exposes performance disparities: mid-tier models succeed in minimal, explicit frameworks (e.g., Flask) but perform substantially worse on average in convention-heavy environments (e.g., FastAPI, Django). Finally, error analysis identifies data-layer defects (e.g., incorrect query composition and ORM runtime violations) as the leading root causes. This work highlights that jointly satisfying functional and structural requirements remains a key open challenge for coding agents.

\end{abstract}

\section{Introduction}

The landscape of software engineering is undergoing a fundamental shift, driven by the increasing use of Large Language Model (LLM) agents capable of autonomously generating code \citep{wang2025openhands,minisweagent2025,yangswe}. These agents have demonstrated remarkable proficiency in "shallow" generation tasks, where objectives are broad and requirements are loosely defined \citep{qwen35blog,openai2026gpt54,hong2024metagpt,KimiK25}.

This flexibility is well suited to a range of development scenarios, such as rapid prototyping, demo applications, frontend interfaces, 
and proof-of-concept implementations, where structural decisions can be left to the agent. 
However, this permissiveness becomes a liability 
when the target is a production-grade backend. Such systems must satisfy functional and non-functional requirements by exposing endpoints that adhere to API contracts, conform to architectural patterns, integrate with databases, and operate through designated ORM layers. The impact of these constraints on agent performance remains largely unexplored.

Recent benchmarks evaluate LLM agents on greenfield code generation (build an application from scratch) or GitHub Issue resolution, yet none adequately captures the challenges of constrained multi-file backend development. Existing efforts either focus on resolving specific issues given a description and existing codebases~\citep{jimenez2024swebench,deng2025swe}, reward unconstrained generation from under-specified natural language (NL) prompts~\citep{luo2026rpg}, directly target single-file solutions via prompting~\citep{vero2025baxbench}, or provide skeleton code for agents to complete~\citep{zhao2025commit0}, overlooking the effect of systematically varying the degree of structural constraint imposed on the agent. 

\begin{figure}[t]
    \centering
    \includegraphics[width=0.93\linewidth]{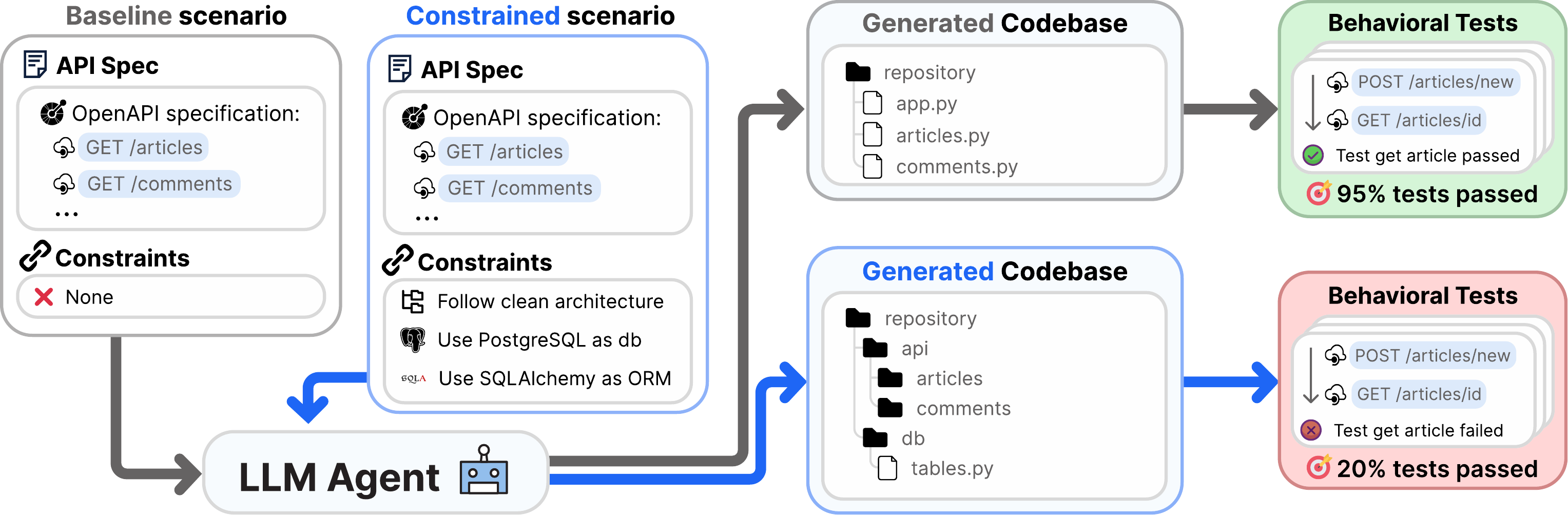}
    \caption{
    Given the same API specification, an LLM agent produces a functional codebase under no constraints (top), 
    but performance degrades with structural constraints 
    (bottom). 
    }
    \label{fig:overview}
\end{figure}

To investigate this scenario, we conduct a systematic study to assess LLM agents in constrained backend generation tasks, leveraging OpenAPI specifications as a structured way of expressing functional requirements. Our design fixes the API contract and deploys a shared suite of end-to-end behavioral tests across all conditions, isolating the effect of \textit{four non-functional constraint dimensions}: framework choice, architectural pattern, database backend, and ORM integration. By systematically layering these constraints, we measure performance degradation as task complexity increases, from baseline generation to fully specified systems requiring an architecture with a prescribed database and ORM layer.

Our empirical findings reveal that 
as the density of non-functional constraints increases, agent performance exhibits a decline. Figure~\ref{fig:overview} illustrates this phenomenon of constraint decay: agents that easily generate functional codebases under baseline conditions fail sharply when strict architectural and data-layer rules are enforced. We demonstrate that even state-of-the-art agents struggle to maintain functional compliance when unconstrained generation is replaced by rigorous adherence to structured specifications.

In this paper, we report the following contributions:
\begin{itemize}[leftmargin=*,noitemsep, topsep=0pt]
    \item \textbf{Evaluation Methodology.} An evaluation pipeline based on OpenAPI specifications and 
    behavioral tests that decouples functional correctness from structural compliance. We open-source 
    80 greenfield generation tasks and 20 feature-implementation tasks.
    
    \item \textbf{Constraint Decay (RQ1).} The empirical identification of \textit{constraint decay}: as explicit structural requirements (architecture, database, ORMs) accumulate, agent performance degrades, with evaluated configurations losing an average of 27.28 points in assertion pass rates (\texttt{A\%}).
    
    \item \textbf{Framework sensitivity (RQ2).} A framework analysis revealing significant disparities in agent success under identical API contracts. Mid-tier models do better in lightweight, explicit frameworks (e.g., Flask) 
    than in convention-heavy 
    environments (e.g., FastAPI); on the strongest models the gap largely closes.

    \item \textbf{Root Cause Taxonomy (RQ3).} An extensive trajectory and error analysis across four models revealing that data-layer defects, specifically incorrect query composition and ORM runtime violations, are the leading root causes for every model, driving $\sim$48\% of agent logic failures on average.
\end{itemize}

\noindent We fully open-source the evaluation pipeline, task suite, agent trajectories, and analysis scripts at \url{https://github.com/francdente/constraint-decay}.

\section{Related Work}
\textbf{Greenfield app generation.} Recent benchmarks have begun to evaluate LLM agents on generating entire apps from scratch, yet none adequately captures the challenges of constrained multi-file, repository-level backend development. 
\cite{luo2026rpg} measure output against existing test suites from well-known libraries but reward unconstrained generation from under-specified prompts. 
\cite{seo2026papercode} target repository generation for reproducing research papers, while 
\cite{zhao2025commit0} provide skeleton libraries for agents to complete. 
\cite{ding2025nl2repo} evaluate coding agents on repository generation from reverse-engineered natural-language specifications of Python codebases. Neither targets backend systems nor systematically varies the degree of structural constraint. 
\cite{vero2025baxbench} is closer to our setting, tasking LLMs with generating backend services from OpenAPI specifications across multiple frameworks and languages, with evaluation through end-to-end functional and security tests. However, it targets single-file solutions via direct prompting rather than multi-file, repositories generated through agentic interaction, and does not study how layering architectural constraints affects generation quality.\\
Our work fixes a single API contract and systematically layers structural constraints across eight frameworks, evaluating implementations against a shared suite of end-to-end behavioral tests fully decoupled from internal code structure, enabling fair comparison regardless of how the agent organizes its solution.

\textbf{Benchmarks for repository-level coding tasks.} 
Issue-based datasets, such as SWE-Bench~\citep{jimenez2024swebench}, offer realistic software engineering environments, yet current agents are approaching saturation on this benchmark. Extensions 
broaden the scope through multi-language support, improved resistance to overfitting, and scalable environment construction~\citep{deng2025swe,rashid2025swe,zan2025multi,jain2025regym}, while complementary benchmarks 
evaluate the ability of agents to add new functionalities \citep{li2025fea,chen2025featbench,miserendino2025swe}. Other works focus on repository-level function or line completion~\citep{liurepobench,ding2023crosscodeeval,zhang2023repocoder} or scalable sandbox-based evaluation~\cite{xie2025repost}.
\\
In our work, we task agents with generating a backend system from scratch, rather than modifying 
an existing codebase, while systematically  increasing structural constraints. 

\textbf{Software Engineering Agents.}
SWE-bench has catalyzed the development of agent architectures. OpenHands provides a framework with tools for file I/O, code execution, and version control in a ReAct-style planner for multi-step reasoning~\citep{wang2025openhands}. SWE-Agent emphasizes configurable agent-computer interfaces 
for interacting with isolated environments~\citep{yangswe}; its variant, Mini-SWE-Agent, reduces the scaffold to 
100 lines of Python with  bash commands. 
\cite{xia2024agentless} forgo autonomous planning  and decompose issue resolution into a fixed 
pipeline: localization, patch generation, and validation.
Our study leverages two of these open-source architectures (Mini-SWE-Agent and OpenHands) to measure how current agents interact with structural constraint density.

\section{Methodology}
To isolate and evaluate the impact of structural constraints on code generation, 
we propose an evaluation methodology inspired by Behavior-Driven Development
~\citep{north2006introducing}. The  principle of our approach is to hold the functional specification constant while systematically varying the non-functional requirements (i.e., structural constraints). This experimental design isolates the effect of the constraints from the logical complexity of the task, thus enabling controlled, equitable comparisons across agents and models.

\subsection{Generation Task Design}
\label{sec:task_design}
We define a greenfield \textit{generation task}: given a comprehensive NL description of a target service, an LLM agent must synthesize a complete, specification-compliant REST API from scratch, i.e., from an empty repository. To execute this task, the agent is provided with an input prompt with four components: (i) the API specification, (ii) the structural constraints to be enforced, (iii) the mandatory files required to execute the generated code, and (iv) a description of the pipeline used to assess the submission (full example in Appendix \ref{app:task_design}).

\textbf{API specification.} We adopt the OpenAPI 3.0 specification\footnote{\url{https://spec.openapis.org/oas/v3.0.0.html}} as the standardized functional requirement format for all tasks. OpenAPI is a machine-readable, language-agnostic interface description standard widely used in industry for API-first development. This structured format yields two  methodological advantages over NL descriptions. First, it removes specification ambiguity: every functional behavior, endpoint signature, request body, response schema, and status code is formally defined. Second, it facilitates the preemptive definition of a behavioral test suite for evaluating any compliant implementation, irrespective of the underlying language, framework, or internal architecture. This approach guarantees a level of evaluation flexibility and behavioral coverage that traditional unit-test-based pipelines cannot achieve, as unit tests are inherently coupled to specific implementation details.

Our specification is derived from the RealWorld Conduit API\footnote{\url{https://realworld-docs.netlify.app/}}. The specification encompasses 19 standard Create, Read, Update, and Delete (CRUD) operations distributed across five resource groups (articles, comments, users, profiles, and tags). The choice of this open-source specification is deliberate. First, it defines operations at a highly granular level while remaining within the context window limits of LLMs. Moreover, the relative simplicity of the functional requests, combined with the models' probable prior exposure to the specification during pre-training, ensures a precise disentanglement between the impact of structural constraints and baseline functional complexity.
\newline We intentionally fix a single API contract to maximize internal validity: holding the functional target constant allows us to isolate the effect of structural constraints, framework, and agent scaffold without conflating them with differences in task semantics.

\textbf{Framework coverage.} Framework is treated as an experimental factor with eight discrete levels, allowing us to test whether the same functional specification exhibits framework-dependent difficulty under identical evaluation conditions. We selected popular frameworks across two of the most widespread language runtimes: Python 3.12 (Flask, FastAPI, Django, aiohttp) and Node.js 20 (Express, Fastify, Hono, Koa). Both runtimes are built upon dynamic and multi-paradigm programming languages for 
 maximal 
 freedom during code generation.

\subsection{Constraint Dimensions}
\label{sec:constraints}

To systematically evaluate the effect of constraints, we define three orthogonal non-functional axes, which can be either included in or omitted from the code specification. 

\textbf{Architectural pattern}. The code architecture is fixed: it must be organized following 
the Clean Architecture pattern~\citep{10.5555/3175742}. Rather than spontaneously define a monolithic or arbitrary file structure, the agent must split the codebase into four domain layers with strict top-down dependency direction: routes/handlers, services/use cases, models/entities, and repository/data access. 
Each layer must reside in its own directory.

\textbf{Database backend}. The database layer is fixed: PostgreSQL or SQLite. The agent must persist data using a specific database engine. 
For PostgreSQL tasks, a pre-configured instance is provided in the execution environment; 
for SQLite, no external service is needed. In both cases, the schema must be created automatically at server startup.

\textbf{ORM integration}. The Object-Relational Mapper (ORM) framework is fixed: SQLAlchemy for Python, Sequelize for Node. The agent must adhere to the idiomatic query logic of the prescribed framework, instead of falling back to raw SQL or alternative mapping.

\begin{table}
\centering
\small
\newcommand{\sep}{\ensuremath{\,\vert\,}}
\begin{tabular}{@{}clc@{}}
\toprule
\textbf{Level} & \textbf{Active Constraints} & \textbf{\# Variants} \\
\midrule
L0 & WF only (baseline) & 8 \\
L1 & WF{+}Arch \sep WF{+}SQLite \sep WF{+}PG & 24 \\
L2 & WF{+}Arch{+}SQLite \sep WF{+}Arch{+}PG \sep WF{+}SQLite{+}ORM \sep WF{+}PG{+}ORM & 32 \\
L3 & WF{+}Arch{+}SQLite{+}ORM \sep WF{+}Arch{+}PG{+}ORM & 16 \\
\bottomrule
\end{tabular}
\vspace{-1ex}
\caption{Constraint levels and task variants. WF = Web Framework (always active), Arch = Clean Architecture, PG = PostgreSQL. Variants within a level are separated by~$\vert$.}
\label{tab:constraint_levels}
\end{table}

Unlike the three axes above, the web framework is always specified: an agent must target some framework to produce executable code. We therefore treat framework as a fixed baseline constraint present at every level. 
To quantify the effect of additional structural complexity, we systematically compose the three axes into task variants of increasing specification density (L0 to L3), as  in Table~\ref{tab:constraint_levels}.
Each variant is instantiated across all 8 frameworks, yielding 80 generation tasks in total.

Although simple and helpful for human developers, these non-functional requirements isolate constraint composition as an added challenge for LLM agents.

\subsection{Evaluation Pipeline}
\label{sec:evaluation}

To guarantee reproducibility and preclude state contamination, each task is executed in an isolated Docker environment following a 
\textit{build} and \textit{evaluate} pipeline. In the build phase, the agent operates within a first  container, where it can code, build, and start the server. Next, in the evaluation phase, the agent code patch is applied in a pristine container and the behavioral test suite is executed. In both phases, a container-dedicated, ephemeral PostgreSQL~16 instance is available (full infrastructure details in Appendix~\ref{app:pipeline}). 

\textbf{Behavioral testing.} Since all task instances target a unified OpenAPI specification, all implementations are evaluated against a unified HTTP behavioral test suite covering the whole specification.
The suite comprises 32 requests covering all 19~API endpoints and 291~assertions that verify response structure, types, status codes, and state transitions across a stateful sequence of interdependent CRUD operations (breakdown in Appendix~\ref{app:test_suite}). This is a methodological choice: by testing the API behavior rather than its implementation, we decouple functional evaluation from internal code structure.

\textbf{Verifier functions.} To separate functional correctness from structural compliance, we introduce orthogonal verifier functions that detect whether the required architecture, database, and ORM constraints are actually satisfied (details in Appendix \ref{app:verifiers}). (i)~\textbf{Architecture verifier}: assesses the compliance to the four canonical Clean Architecture layers (routes, services, repositories, models) and scans every import to verify dependency order. (ii)~\textbf{Database verifier}: scans the source lines for usage evidence of the imposed database engine while checking that no 
alternative engines are used; lock files are excluded. 
(iii)~\textbf{ORM verifier}: detects whether the designated Object-Relational Mapper is used rather than raw SQL or framework-native data layer.

We define task success as the intersection of behavioral validity and constraint compliance. A completed task 
passes the entire behavioral test suite and satisfies each applicable verifier.

\section{Experimental Setup}

\paragraph{Metrics.}
\label{sec:metrics}

We define our primary evaluation axes as follows. \textbf{Assert\%} (A\%): the mean fraction of behavioral test assertions passed across all runs of a given configuration. It captures partial progress, distinguishing configurations that slightly miss full compliance from largely non-functional implementations. We report A\% as the primary metric, since the all-or-nothing nature of \texttt{pass@1} over the entire test suite amplifies noise. \textbf{Pass@k}  \citep[using the unbiased estimator of][]{chen2021evaluating}: 
the probability that at least one of $k$ independent samples satisfies the complete set of behavioral and structural constraints. 

Each task is executed for $n{=}3$ independent trials per agent-model configuration. 

\paragraph{Agents and Models.}
\label{sec:agents}
We evaluate two popular and open-source agent architectures 
of different tooling complexity. 
(i)~\textbf{Mini-SWE-Agent}: a minimal scaffold (${\sim}$100 lines of Python) that interacts with the environment exclusively through bash commands~\citep{minisweagent2025}. Despite its simplicity, it achieves strong performance on SWE-Bench with frontier LLMs ($>$70\%). 
(ii)~\textbf{OpenHands}: a full-featured agent framework providing dedicated tools for file editing, terminal execution, code search, and task tracking~\citep{wang2025openhands}.

To ensure experimental validity and calibrate the agents for code generation, we apply systematic adjustments to the scaffold configuration through prompt refinement and pipeline adjustments. We inject a re-aligned system prompt into Mini-SWE-Agent to steer its execution trace toward a generative pipeline. Additionally, to mitigate premature halting behaviors \citep{ding2025nl2repo} and enforce complete evaluation trajectories, we introduce a termination criterion for OpenHands 
(details in Appendix \ref{app:agent_scaffolds}). For both agents, we impose a highly permissive upper bound on the maximum number of iterations (300 for Mini-SWE-Agent and 200 for OpenHands) while leaving the cost limit unbounded. 

We pair each scaffold with models spanning a broad capability range. We distinguish four tiers. (i)~\textbf{Small open agentic models}: compact code-specialists, i.e., Devstral-Small~(24B)~\citep{rastogi2025devstralfinetuninglanguagemodels} and Qwen3-Coder-Next~(80B)~\citep{qwen_qwen3_coder_next_tech_report}. 
(ii)~\textbf{Large open instruct model}: general purpose architectures, i.e., Qwen3-235B-Instruct~(235B total\,/\,22B active)~\citep{qwen3}. 
(iii)~\textbf{Large open agentic models}: SOTA open-weight models, specifically MiniMax-M2.5~\citep{MiniMax} and Kimi-K2.5~\citep{KimiK25}. 
(iv)~\textbf{Closed models}: frontier proprietary engines, i.e., GPT-5-mini~\citep{openai2025gpt52}, GPT-5.2~\citep{openai2025gpt52}, and GPT-5.4~\citep{openai2026gpt54}. 
The larger models 
rank among the top 15 on the SWE-Bench Verified leaderboard.

\paragraph{Cost-driven evaluation scope.} OpenHands's richer tool set leads to substantially larger context windows than Mini-SWE-Agent's scaffold: 12.9$\times$ more tokens for GPT-5-mini.
To reduce cost for generation tasks we restrict Kimi-K2.5 and GPT-5.2 to Mini-SWE-Agent only. Moreover, as in~\cite{zhao2025commit0}, for these two LLMs and the OpenHands + MiniMax-M2.5 pair we use a \textit{representative task subset} (marked * in Table~\ref{tab:main_results}). 
The subset spans two frameworks per runtime: aiohttp and FastAPI (Python), Express and Fastify (Node), and one variant per constraint level: L0, L1 (Clean Architecture), L2 (Clean Architecture\,+\,PostgreSQL), L3 (Clean Architecture\,+\,PostgreSQL\,+\,ORM), yielding 16~tasks and 48~runs per model-agent pair.
Across the full study, our evaluation used $\sim$4~billion tokens 
(details in Appendix \ref{app:token_consumption}).

\section{Results}
We organize the analysis around three research questions.\\
\textbf{RQ1.} How does the accumulation of structural constraints affect LLM agent performance (\S\ref{sec:constraint_decay})?
\textbf{RQ2.} Does the choice of web framework significantly influence agent success under the same API contract (\S\ref{sec:cross_framework})?
\textbf{RQ3.} What are the primary root causes of agent failure in constrained backend generation (\S\ref{sec:failure_analysis})?

\subsection{RQ1: Effect of Constraint Accumulation}
\label{sec:constraint_decay}


Table~\ref{tab:main_results} summarizes the main results. Each cell reports assertion pass rate A\% and \texttt{pass@1}~(\%), computed per-task and averaged across tasks. Three model-agent pairs (OpenHands + MiniMax-M2.5, Mini-SWE + Kimi-K2.5, Mini-SWE + GPT-5.2) were evaluated on the 16-task subset due to cost constraints. To validate this choice, we re-evaluated the eight full-set configurations (3 models $\times$ 2 agents, and 2 models on Mini-SWE only, excluding Devstral-Small) on the same 16 tasks: A\% scores correlate at Pearson $r = 0.94$ and Spearman $\rho = 0.93$ ($p < 10^{-14}$, $N = 32$).
Specifically, we compared each full-set model's A\% on all 80 tasks against its A\% on the same 16-task subset, obtaining 32 paired observations (8 model--agent configurations $\times$ 4 constraint levels) that show near-perfect agreement in both absolute scores and ranking order (details in Appendix~\ref{app:subset_validity}).

\begin{table}
\centering
\small
\begin{tabular}{@{}ll cccc cccc r@{}}
\toprule
& & \multicolumn{4}{c}{Assert\% (A\%)} & \multicolumn{4}{c}{\texttt{pass@1}} & \\
\cmidrule(lr){3-6} \cmidrule(lr){7-10}
\textbf{Agent} & \textbf{Model} & \textbf{L0} & \textbf{L1} & \textbf{L2} & \textbf{L3} & \textbf{L0} & \textbf{L1} & \textbf{L2} & \textbf{L3} & \textbf{$\Delta$A\%} \\
\midrule
Mini-SWE & GPT-5-mini & 51.7 & 46.8 & 27.1 & 23.7 & 25.0 & 22.2 & 16.7 & 4.2 & -28.0 \\
OpenHands & GPT-5-mini & 65.8 & 63.2 & 56.7 & 52.2 & 62.5 & 47.2 & 39.6 & 33.3 & -13.6 \\
Mini-SWE & Qwen3-Coder-Next & 86.4 & 66.4 & 52.6 & 46.1 & 45.8 & 25.0 & 13.5 & 6.2 & -40.2 \\
OpenHands & Qwen3-Coder-Next & 73.0 & 51.7 & 42.7 & 27.6 & 45.8 & 16.7 & 16.7 & 4.2 & -45.5 \\
Mini-SWE & Qwen3-235B-A22B & 29.6 & 10.7 & 9.1 & 2.3 & 20.8 & 1.4 & 2.1 & 0.0 & -27.3 \\
OpenHands & Qwen3-235B-A22B & 26.2 & 17.7 & 3.1 & 0.8 & 12.5 & 2.8 & 0.0 & 0.0 & -25.4 \\
\midrule
Mini-SWE & MiniMax-M2.5 & 90.5 & 79.3 & 75.0 & 66.8 & 62.5 & 41.7 & 37.5 & 29.2 & -23.7 \\
Mini-SWE & GPT-5.4 & 97.5 & 81.9 & 81.9 & 80.0 & 62.5 & 53.6 & 50.5 & 64.6 & -17.5 \\
\midrule
OpenHands & MiniMax-M2.5* & 95.6 & 97.0 & 87.3 & 78.6 & 41.7 & 41.7 & 16.7 & 8.3 & -17.0 \\
Mini-SWE & Kimi-K2.5* & 85.4 & 70.9 & 62.9 & 53.7 & 66.7 & 41.7 & 8.3 & 33.3 & -31.7 \\
Mini-SWE & GPT-5.2* & 78.2 & 49.3 & 27.1 & 48.0 & 58.3 & 33.3 & 0.0 & 25.0 & -30.2 \\
\bottomrule
\end{tabular}
\caption{Assertion pass rate A\% and \texttt{pass@1} across constraint levels. L0 = baseline; L1--L3 impose increasingly specific structural requirements. 
* = evaluated on task subset.}
\label{tab:main_results}
\end{table}

Without structural constraints (L0), agents perform well: Qwen3-Coder-Next,
MiniMax-M2.5, GPT-5.4, and Kimi-K2.5 exceed 85\% A\%, confirming
that 
compact code-specialists such as
Qwen3-Coder-Next handle end-to-end backend generation when
given full architectural freedom. Smaller agentic models
and non-specialized ones 
struggle even at L0. 
Given its near-zero Assert \% (see Table \ref{tab:verifier_comparison}), we exclude Devstral-Small from the remainder of the discussion.

\textbf{Overall decay.}
Structural constraints cause steep, universal performance degradation.
Restricting to the nine capable configurations (L0
A\%~$>$~50\%) in Table~\ref{tab:main_results}, A\% drops by an
average of 27.48~percentage points from L0 to L3: a relative loss of more than
30\% of baseline performance. The worst case,
OpenHands\,+\,Qwen3-Coder-Next, loses 45~pp (62\% of its L0 score), while the most resilient configuration, OpenHands\,+\,MiniMax-M2.5,
drops 17~pp. 

The gap between A\% and \texttt{pass@1} is consistently large: even the strongest L3 configuration (OpenHands\,+\,MiniMax-M2.5) reaches 78.6\% A\% but only 8.3\% \texttt{pass@1}.
Because a single assertion failure among the entire suite zeroes out a run, \texttt{pass@1} is noisy for evaluating generation tasks; A\% captures partial progress more stably and is therefore the primary metric to look at.
This also underscores how LLM agents still lack the cross-file consistency and constraint adherence needed for deployment without manual intervention.

The variance of the drop across levels is substantial, so the precise effect size should be interpreted with caution. However, the central qualitative finding is robust: adding structural constraints consistently worsens performance.

\textbf{Scaffold effect.}
The choice of agent scaffold has a substantial impact. OpenHands
outperforms Mini-SWE-Agent for GPT-5-mini and MiniMax-M2.5 yet
underperforms for Qwen3-Coder-Next; however, the constraint-decay trend is
consistent across both scaffolds.

\begin{table}
\centering
\small
\begin{subtable}[t]{0.48\linewidth}
\centering
\begin{tabular}{@{}l r@{$\;\pm\;$}l@{}}
\toprule
\textbf{Constraint} & \multicolumn{2}{c}{\textbf{Avg $\Delta$ (pp)}} \\
\midrule
Clean architecture   & $-$8.8 & 1.7  \\
PostgreSQL           & $-$20.4 & 2.5  \\
SQLite               & $-$16.1 & 2.6  \\
SQLAlchemy           & $-$4.8 & 2.6  \\
Sequelize            & $+$1.8 & 2.2  \\
\bottomrule
\end{tabular}
\caption{Marginal effect of each constraint on A\%, via matched-pair differences ($\pm$ stderr of the mean).}
\label{tab:marginal_constraint}
\end{subtable}
\hfill
\begin{subtable}[t]{0.48\linewidth}
\centering
\begin{tabular}{@{}l cc@{}}
\toprule
& \multicolumn{2}{c}{\textbf{pass@1 (\%)}} \\
\cmidrule(lr){2-3}
\textbf{Model} & \textbf{Mini-SWE} & \textbf{OpenHands} \\
\midrule
GPT-5-mini      & 15.0 & 48.3 \\
GPT-5.2$^\dagger$ & 50.0   & 55.0 \\
MiniMax-M2.5    & 46.7 & 38.3 \\
Qwen3-Coder-Next     & 16.7 & 18.3 \\
\bottomrule
\end{tabular}
\caption{\texttt{pass@1}~(\%) on feature implementation tasks. $^\dagger$Single run instead of 3 for GPT-5.2 only.}
\label{tab:feature_results}
\end{subtable}
\caption{(a)~Marginal constraint effects. (b)~Feature implementation results.}
\label{tab:constraint_and_feature}
\end{table}

\textbf{Per-constraint analysis.} To isolate the marginal effect of each constraint, we adopt a matched-pair design: for every constraint~$c$, we identify all task pairs that differ by exactly~$c$ while keeping framework and all other constraints identical, then compute $\Delta(\text{A\%}) = \text{A\%}_{\text{with}\,c} - \text{A\%}_{\text{without}\,c}$ and average across all pairs and model--agent configurations. Table~\ref{tab:marginal_constraint} reports the results. 
Specifying a database engine (which is fundamental for a production-ready backend) is by far the most impactful constraint. 
Clean architecture also adds a significant 
penalty, consistent with the overhead of enforcing layered separation. 
ORM's effects are smaller and, for GPT-5.2, the ORM constraint even reduces ambiguity by making the intended data-access pattern more explicit.

\textbf{Verifier impact on A\%.} All A\% scores in Table~\ref{tab:main_results} incorporate structural compliance: when a run violates any applicable verifier, its assertion score is set to zero before averaging. To quantify the impact of this choice, we also compute A\% without verifier enforcement. The two variants differ by at most 2.7~pp on any single configuration, and the average L0$\to$L3 drop changes only from 26 to 27~pp. This confirms that the constraint-decay signal is driven by genuine functional failures rather than verifier artifacts (full comparison in Appendix~\ref{app:verifier_impact}).

\textbf{Sanity check: feature implementation tasks.}
A natural concern is whether constraint decay is an artifact of the synthetic greenfield setting.
To test this, we construct 20 feature implementation tasks by ablating feature groups (161--2{,}604 lines across 2--46 files) from community-maintained RealWorld repositories (actual implementations of the API contract used in the generation tasks) that already embed a layered architecture, PostgreSQL, and an ORM, effectively simulating an L3 condition with implicit constraints. Each agent must read the existing codebase, infer its conventions, and re-implement the missing functionality (full task design in Appendix~\ref{app:feature_task_design}). As shown in Table~\ref{tab:feature_results}, performance remains low: only GPT-5.2 exceeds 50\% \texttt{pass@1}. While we cannot isolate structural constraints as the sole cause of difficulty in this setting, the results confirm that the challenges observed in greenfield generation are not artifacts of asking agents to build repositories from scratch; they persist when agents must infer and respect constraints from an existing codebase.

\noindent\fbox{%
    \parbox{\linewidth}{\textit{Takeaways for RQ1}. Structural constraints cause steep, universal decay: capable configurations 
    lose 27.48~pp A\% from L0 to L3 on average. 
    Database constraints drive most of this decline. 
}}
\vspace{-1ex}

\subsection{RQ2: Framework Sensitivity}
\label{sec:cross_framework}

\begin{figure}[h]
    \centering
    \includegraphics[width=0.95\linewidth]{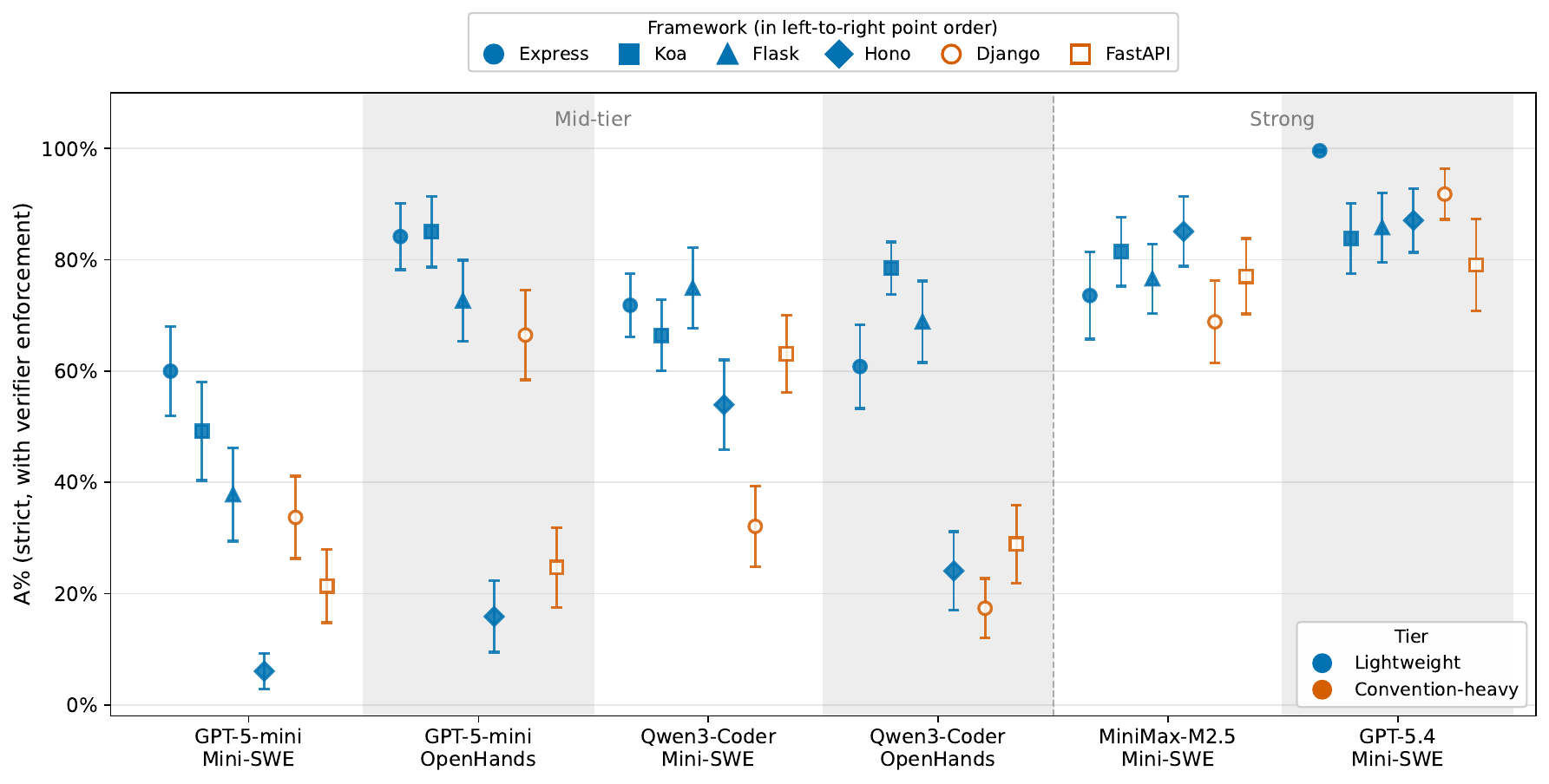}
    \caption{A\% by framework across model--scaffold configurations, aggregated over constraint levels with $\pm1$ standard error across individual runs.}
    \label{fig:framework_sensitivity}
\end{figure}

Since framework is the one constraint present at every level, a natural question is how much the choice of framework matters.
All eight frameworks implement the same API contract and share the same test suite, enabling a controlled comparison.
Figure~\ref{fig:framework_sensitivity} reports A\% by framework, aggregated across constraint levels, for six model--scaffold configurations: four mid-tier ones (GPT-5-mini and Qwen3-Coder-Next, each with both scaffolds) and two strong ones (MiniMax-M2.5 and GPT-5.4, both with Mini-SWE-Agent).
The figure shows the top-3 and bottom-3 frameworks by average A\% (aiohttp and Fastify, which rank fourth and fifth, are omitted for legibility); the complete framework table is in Appendix~\ref{app:framework_leaderboard}.

\textbf{Framework influence.}
On the four mid-tier configurations, the best-to-worst spread is 43--69~pp and the framework ranking is stable: Express, Koa and Flask occupy the top three positions in all four (avg.\ 69.2\%, 69.8\%, 63.6\% A\%), Django, FastAPI and Hono the bottom three (37.4\%, 34.5\%, 25.0\%). Only the ordering within each tier shifts: Koa and Express lead with OpenHands\,+\,GPT-5-mini, 
Flask with Mini-SWE\,+\,Qwen3-Coder-Next. 
The top tier shares a minimal, explicit API surface with no implicit conventions, whereas
Django and FastAPI 
are penalized by implicit configuration: Django's convention-driven structure and auto-discovery, FastAPI's type-hint-driven validation. Hono trails despite a comparable API surface to Express, likely because 
Hono targets edge runtimes and requires a compatibility adapter on Node.js (the runtime used in our study), a setup step that may be underrepresented in training data.
The strong configurations break this pattern. The spread falls to 16~pp for MiniMax-M2.5 and 21~pp for GPT-5.4, the ranking does not hold anymore, and most inter-framework differences fall within $\pm$stderr.

\noindent\fbox{%
    \parbox{\linewidth}{\textit{Takeaways for RQ2.} Framework choice is a major sensitivity axis for mid-tier configurations: lightweight frameworks lead convention-heavy ones by 26--45~pp on average.
}}

\subsection{RQ3: Root Causes of Failure}
\label{sec:failure_analysis}

We perform a 
failure analysis on unsuccessful runs for four Mini-SWE-Agent configurations spanning the capability range: Qwen3-Coder-Next (194/240 runs failed), GPT-5-mini (200/240), MiniMax-M2.5 (145/240), and GPT-5.4 (103/240). Following established approaches~\citep{deng2025swe}, we use GPT-5.2 (temperature 0) to classify each failure based on the last 20 trajectory turns, behavioral test results, server logs, and static verifier outputs. We define a two-level taxonomy (Table~\ref{tab:failure_analysis}): seven coarse categories (top), with logic errors only further sub-classified into six root causes (bottom). We validate the judge on a stratified sample of 50 Qwen3-Coder-Next logic errors against manual labels, obtaining Cohen's $\kappa = 0.975$, aided by the largely unambiguous signal from server logs and test output (details in Appendix~\ref{app:judge_validation}).

\begin{table}[h]
\centering
\small
\begin{tabular}{@{}l rrrr@{}}
\toprule
 & \textbf{Qwen3-Coder} & \textbf{GPT-5-mini} & \textbf{MiniMax-M2.5} & \textbf{GPT-5.4} \\
\midrule
\multicolumn{5}{@{}l}{\textit{Coarse failure category} (\% of failed runs)} \\
Logic error               & 70.6 & 31.5 & 76.6 & 62.1 \\
Server startup failure    & 12.4 & 50.5 & 16.6 & 31.1 \\
Incomplete implementation &  9.3 & 12.5 &  2.8 &   -- \\
Schema / format error     &  3.6 &  2.5 &  2.8 &  5.8 \\
Stuck in loop             &  3.1 &   -- &   -- &   -- \\
Constraint violation      &  1.0 &  1.0 &  1.4 &   -- \\
Premature termination     &   -- &  2.0 &   -- &  1.0 \\
\textbf{Total failed runs} & \textbf{194} & \textbf{200} & \textbf{145} & \textbf{103} \\
\midrule
\multicolumn{5}{@{}l}{\textit{Logic-error subcategory} (\% of logic errors)} \\
Incorrect query logic      & 25.5 & 15.9 & 29.7 & 50.0 \\
DB / ORM runtime error     & 21.2 & 20.6 & 19.8 & 10.9 \\
Auth misconfiguration      & 22.6 & 15.9 &  7.2 &  1.6 \\
Framework idiosyncrasy     &  9.5 & 25.4 & 24.3 & 21.9 \\
Business logic defect      & 11.7 & 19.0 &  9.9 & 12.5 \\
State propagation failure  &  9.5 &  3.2 &  9.0 &  3.1 \\
\textbf{Total logic errors} & \textbf{137} & \textbf{63} & \textbf{111} & \textbf{64} \\
\bottomrule
\end{tabular}
\caption{
Failure taxonomy for four models with Mini-SWE-Agent, all on the full task set. Top: coarse categories as \textbf{\%} of failed runs; 
bottom: 
subcategories of \textit{logic errors} only.}
\label{tab:failure_analysis}
\end{table}

\textbf{Coarse profile.}
Logic errors, where the server starts and registers correct routes but behaves incorrectly, are the dominant failure mode for three of the four models (62.1--76.6\%), consistent with the persistent gap between Assert\% and \texttt{pass@1} in \S\ref{sec:constraint_decay}. GPT-5-mini is the exception: half of its failures (50.5\%) are server startup failures, i.e.\ the agent never produces a runnable server, and only 31.5\% reach the logic-error stage. The remaining categories together account for less than 20\% of failures.

\textbf{Logic-error root causes.}
We classify the logic errors 
into six root-cause categories:
\emph{incorrect query logic}: SQL queries execute but return wrong results due to incorrect joins, filters, or dialect-incompatible operators;
\emph{DB/ORM runtime error}: query logic is conceptually correct but crashes from ORM API misuse;  
\emph{auth misconfiguration}: broken token handling or header parsing causes 
401 responses;
\emph{framework idiosyncrasy}: correct application code is blocked by an unhandled framework-specific default;
\emph{business logic defect}: infrastructure operates correctly but domain behavior deviates from the specification;
\emph{state propagation failure}: mutations succeed but their effects are not reflected in subsequent reads.

Data-layer defects (\emph{incorrect query logic} plus \emph{DB/ORM runtime errors}) are the leading root cause for all four models, accounting for 36.5--60.9\% of logic errors (48.4\% on average). Together, these data-layer failures confirm that database constraints impose the steepest penalty.
\emph{Framework idiosyncrasy} is the second recurring theme (9.5--25.4\%) and is overwhelmingly one framework: 56 of the 70 cases across all four models are Fastify's rejection of empty-body POST requests on the favorite and follow endpoints.
\emph{Auth misconfiguration} declines monotonically with capability (22.6\%, 15.9\%, 7.2\%, 1.6\%), suggesting that token-prefix and header-parsing conventions are the kind of knowledge stronger models reliably absorb.
The prevalence of data-layer defects does not contradict the low marginal effect of ORM constraints observed in Table \ref{tab:marginal_constraint}.
Because the marginal effect of ORMs is computed relative to a baseline that already includes a database, it merely measures the cost of forcing an ORM in place of raw SQL—not the cost of data-layer interaction itself. The underlying difficulty of correctly interfacing with the data layer is high in both scenarios; it simply manifests as ORM runtime errors when the ORM constraint is active, and as raw SQL errors when it is not.

\noindent\fbox{%
    \parbox{\linewidth}{\textit{Takeaways for RQ3.} Logic errors dominate failures for three of the four models analyzed (62--77\%). Within logic errors, data-layer defects are the leading root cause for every model.
}}

\section{Conclusion}
Our systematic study exposes a phenomenon of \textit{constraint decay} in LLM-based coding agents. While current models excel at unconstrained generation, their performance drops when forced to navigate explicit architectural rules. For end-users, this dichotomy implies that agents are reliable for rapid prototyping but remain unreliable for production-grade backend development. Overcoming this bottleneck requires a shift for agent developers: moving beyond purely functional benchmarks to actively integrate structural awareness, potentially through retrieval-augmented framework documentation, constraint-oriented planning, or targeted pre-training on convention-heavy codebases.

\section*{Acknowledgement}
This work was partially supported by the AI4SWEng project, funded by the European Union’s Horizon Europe research and innovation programme under grant agreement No. 101189909, and by the 3IA Côte d’Azur Investments in the IA-cluster project managed by the National Research Agency (ANR) with the reference number ANR-23-IACL-0001.

\bibliography{colm2026_conference}
\bibliographystyle{colm2026_conference}

\appendix
\vspace{2ex}
\noindent {\large\bfseries APPENDIX}

\section{Subset Representativeness}
\label{app:subset_validity}

Due to cost constraints, three model-agent pairs (OpenHands + MiniMax-M2.5, Mini-SWE + Kimi-K2.5, Mini-SWE + GPT-5.2) were evaluated only on a 16-task subset comprising 4 frameworks (aiohttp, Express, FastAPI, Fastify) $\times$ 4 constraint levels (L0--L3), where each framework follows the pipeline: unconstrained $\to$ +clean\_architecture $\to$ +postgres $\to$ +postgres+ORM.

To verify that results on this subset are representative of the full 80-task benchmark, we re-evaluated the eight full-set configurations (GPT-5-mini, Qwen3-Coder-Next, Qwen3-235B-A22B, each with both Mini-SWE-Agent and OpenHands, while MiniMax-M2.5 and GPT-5.4 with Mini-SWE only; Devstral-Small excluded due to near-zero scores) on the same 16 tasks and compared against their full-set scores. This yields $N = 32$ paired observations (8 configurations $\times$ 4 constraint levels), where each pair consists of (full-set A\%, subset A\%) for the same configuration and level.

\textbf{Assert\%} scores on the full set and subset correlate at Pearson $r = 0.948$ ($p < 10^{-15}$) and Spearman $\rho = 0.932$ ($p < 10^{-14}$). Critically, cross-model rankings are preserved almost perfectly at every constraint level. The L0$\to$L3 decay direction is also consistent across all configurations: every configuration that shows performance degradation on the full set also degrades on the subset.

These results confirm that the 16-task subset reliably preserves both absolute performance rankings and constraint-decay trends, supporting the validity of subset-only evaluations for the cost-constrained models reported in Table~\ref{tab:main_results}.

\begin{table}[h]
\centering
\footnotesize
\setlength{\tabcolsep}{3.2pt}
\begin{tabular}{@{}ll rrrr rrrr rrrr@{}}
\toprule
& & \multicolumn{4}{c}{Full-set A\%} & \multicolumn{4}{c}{Subset A\%} & \multicolumn{4}{c}{$\Delta$ (pp)} \\
\cmidrule(lr){3-6} \cmidrule(lr){7-10} \cmidrule(lr){11-14}
\textbf{Agent} & \textbf{Model} & \textbf{L0} & \textbf{L1} & \textbf{L2} & \textbf{L3} & \textbf{L0} & \textbf{L1} & \textbf{L2} & \textbf{L3} & \textbf{L0} & \textbf{L1} & \textbf{L2} & \textbf{L3} \\
\midrule
Mini-SWE & GPT-5-mini & 51.7 & 46.8 & 27.1 & 23.7 & 67.6 & 47.0 & 8.0 & 32.5 & \llap{$-$}15.9 & \llap{$-$}0.2 & \llap{$+$}19.2 & \llap{$-$}8.8 \\
OpenHands & GPT-5-mini & 65.8 & 63.2 & 56.7 & 52.2 & 81.6 & 70.4 & 41.1 & 33.8 & \llap{$-$}15.8 & \llap{$-$}7.3 & \llap{$+$}15.5 & \llap{$+$}18.3 \\
Mini-SWE & Qwen3-Coder & 86.4 & 66.4 & 52.6 & 46.1 & 87.7 & 67.9 & 45.3 & 51.2 & \llap{$-$}1.4 & \llap{$-$}1.5 & \llap{$+$}7.3 & \llap{$-$}5.1 \\
OpenHands & Qwen3-Coder & 73.0 & 51.7 & 42.7 & 27.6 & 76.7 & 59.5 & 21.4 & 14.7 & \llap{$-$}3.6 & \llap{$-$}7.8 & \llap{$+$}21.2 & \llap{$+$}12.8 \\
Mini-SWE & Qwen3-235B-A22B & 29.6 & 10.7 & 9.1 & 2.3 & 25.9 & 15.9 & 1.8 & 2.8 & \llap{$+$}3.7 & \llap{$-$}5.2 & \llap{$+$}7.3 & \llap{$-$}0.6 \\
OpenHands & Qwen3-235B-A22B & 26.2 & 17.7 & 3.1 & 0.8 & 32.1 & 29.4 & 0.0 & 0.1 & \llap{$-$}5.9 & \llap{$-$}11.7 & \llap{$+$}3.1 & \llap{$+$}0.7 \\
Mini-SWE & MiniMax-M2.5 & 90.5 & 79.3 & 75.0 & 66.8 & 88.6 & 92.5 & 66.8 & 58.3 & \llap{$+$}1.9 & \llap{$-$}13.2 & \llap{$+$}8.2 & \llap{$+$}8.6 \\
Mini-SWE & GPT-5.4 & 97.5 & 81.9 & 81.9 & 80.0 & 96.2 & 80.9 & 87.5 & 74.4 & \llap{$+$}1.3 & \llap{$+$}1.0 & \llap{$-$}5.7 & \llap{$+$}5.6 \\
\midrule
\multicolumn{2}{@{}l}{MAE} & & & & & & & & & 6.2 & 6.0 & 10.9 & 7.6 \\
\bottomrule
\end{tabular}
\caption{Subset representativeness: A\% on the full 80-task set vs.\ the 16-task subset for each of the 8 full-evaluation configurations (Devstral-Small excluded). $\Delta$ = full$-$subset. MAE = mean absolute difference. Pearson $r = 0.948$, Spearman $\rho = 0.932$ ($N = 32$, $p < 10^{-14}$).}
\label{tab:more_results}
\end{table}

\section{LLM-as-Judge Validation}
\label{app:judge_validation}

To assess the reliability of the GPT-5.2 judge used for logic-error sub-categorization (\S\ref{sec:failure_analysis}), we drew a stratified random sample of 50 logic errors from the Qwen3-Coder-Next failure set ($n{=}137$), preserving population proportions across the six subcategories. A human  (one of the authors) independently labeled each entry using the same taxonomy and evidence (last 20 trajectory turns, test-failure summary, server logs), without viewing the judge's label.

Table~\ref{tab:judge_validation} reports the results. Overall accuracy is 98\,\% (49/50) with Cohen's $\kappa = 0.975$. Four of the six categories achieve perfect precision, recall, and F1. The near-perfect agreement is largely attributable to the rich signal in server logs and test outputs, which make the root cause of most logic errors unambiguous.

\begin{table}[h]
\centering
\footnotesize
\begin{tabular}{@{}l rrr r@{}}
\toprule
\textbf{Subcategory} & \textbf{Prec.} & \textbf{Rec.} & \textbf{F1} & \textbf{Support} \\
\midrule
auth\_misconfiguration      & 100.0 & 100.0 & 100.0 & 11 \\
incorrect\_query\_logic      &  92.3 & 100.0 &  96.0 & 12 \\
state\_propagation\_failure  & 100.0 & 100.0 & 100.0 &  5 \\
database\_runtime\_error     & 100.0 & 100.0 & 100.0 & 11 \\
framework\_idiosyncrasy      & 100.0 &  80.0 &  88.9 &  5 \\
business\_logic\_defect      & 100.0 & 100.0 & 100.0 &  6 \\
\midrule
\textbf{Macro avg}           &  98.7 &  96.7 &  97.5 & 50 \\
\bottomrule
\end{tabular}
\caption{Per-category precision, recall, and F1 (\%) of the GPT-5.2 judge against human annotations on the 50-sample validation set. Cohen's $\kappa = 0.975$; overall accuracy = 98\%.}
\label{tab:judge_validation}
\end{table}

\section{Verifier Functions}
\label{app:verifiers}

The three verifier functions introduced in \S\ref{sec:evaluation} operate on the unified diff (patch) produced by each agent run.  They parse only the added lines of the patch (i.e., the source code the agent actually wrote) and apply static pattern matching to determine structural compliance.  No execution or dynamic analysis is performed.

\subsection{Architecture Verifier}

This verifier enforces the Clean Architecture constraint through two checks:

\textbf{Layer presence.} The prompt prescribes four layers, each in its own directory: routes/handlers (rank~3), services/use cases (rank~2), repositories/data access (rank~1), and models/entities (rank~0).  Agents use diverse naming conventions, so the verifier maps each directory component against a set of aliases (e.g., \texttt{routes}, \texttt{handlers}, \texttt{routers}, \texttt{views} all map to the routes layer).  A solution is considered layered if at least 3 of the 4 canonical layers are present as distinct directories.

\textbf{Dependency direction.}  For every source file classified into a layer, the verifier extracts all import statements (Python \texttt{import}/\texttt{from~\ldots~import} and JavaScript \texttt{require()}/\texttt{import~\ldots~from}) and resolves each import path to a target layer using the same alias mapping.  A dependency violation is recorded when a file in a lower-rank layer imports from a higher-rank layer (e.g., a repository importing from a route handler).

A solution is compliant if and only if both checks pass: sufficient layer presence and no upward dependency violations.

\subsection{Database Verifier}

This verifier checks that the agent used the prescribed database engine (SQLite or PostgreSQL) and did not use any alternative.  It scans all added source lines (excluding lock files, which list optional peer dependencies for all supported dialects) for two sets of regex evidence patterns:

\textbf{SQLite evidence.} \texttt{sqlite3}/\texttt{aiosqlite} imports, \texttt{sqlite:///} connection strings, Django's \texttt{sqlite3} backend, Node packages (\texttt{sqlite3}, \texttt{better-sqlite3}), and Sequelize \texttt{dialect:~'sqlite'}.

\textbf{PostgreSQL evidence.} \texttt{psycopg2}/\texttt{asyncpg} imports, \texttt{postgresql://} connection strings, SQLAlchemy's PostgreSQL dialect imports, Django's \texttt{postgresql} backend, Node packages (\texttt{pg}, \texttt{pg-promise}), Sequelize \texttt{dialect:~'postgres'}, and host references to the Docker service name \texttt{postgres}.

A solution is compliant when at least one evidence pattern for the expected database is found and no evidence of the alternative database is present.  A special case handles Django projects that retain the default SQLite backend in \texttt{settings.py} even when PostgreSQL is the active configuration (the second \texttt{DATABASES} assignment overrides the first).

\subsection{ORM Verifier}

This verifier checks for evidence of the prescribed ORM (SQLAlchemy for Python tasks, Sequelize for Node tasks). It scans added lines for:

\textbf{SQLAlchemy.} \texttt{import~sqlalchemy} / \texttt{from~sqlalchemy~import}, or any \texttt{sqlalchemy} mention in dependency files.

\textbf{Sequelize.} \texttt{require('sequelize')} / \texttt{import~\ldots~from~'sequelize'}, \texttt{new~Sequelize()} constructor calls, or \texttt{"sequelize":} entries in \texttt{package.json}.

A solution is compliant if at least one evidence pattern for the expected ORM is found.  Unlike the database verifier, the presence of an alternative ORM is not penalized: some agents import utility libraries that happen to reference other ORMs without actually using them for data access.

\subsection{Limitations}

All three verifiers are deliberately conservative in their design:

\textbf{Static-only.} The verifiers analyze source text, not runtime behavior.  An agent could satisfy the verifier's patterns while violating constraints at runtime (e.g., importing SQLAlchemy but executing raw SQL).  This means false positives are possible, but as discussed in Appendix~\ref{app:verifier_impact}, they can only weaken the observed decay signal.

\textbf{Pattern-based.} Regex matching cannot capture all possible naming conventions or import patterns.  We mitigate this with broad alias sets and case-insensitive matching, and validate empirically that the false-reject rate is $\leq$0.5\% (\S\ref{app:verifier_impact}).

\section{Verifier Impact on Reported Metrics}
\label{app:verifier_impact}

A potential concern with our evaluation is that the static verifier functions (architecture, database, ORM) could introduce measurement artifacts: wrong rejections (incorrectly rejecting compliant runs, zeroing their A\% contribution) would artificially inflate the constraint-decay signal, while missed violations (incorrectly accepting non-compliant runs) would weaken it by allowing structurally incorrect implementations to retain their A\% scores.

To quantify the actual impact, we compute two variants of A\%: (i) with verifier enforcement (Table~\ref{tab:main_results}), where a run that violates any applicable verifier has its assertion score set to zero before averaging; and (ii) without verifier enforcement, using raw behavioral-test scores regardless of structural compliance. Table~\ref{tab:verifier_comparison} reports both variants side-by-side.

\textbf{Wrong rejections.}
The maximum difference between the two A\% variants across all configurations and constraint levels is 2.7~pp (Mini-SWE\,+\,Qwen3-Coder-Next at L3). The average L0$\to$L3 decay changes from 26~pp (without enforcement) to 27~pp (with enforcement). This near-identity confirms that verifier wrong rejections have negligible impact: the overwhelming majority of test-passing runs also satisfy all structural constraints, and the few that do not (${\leq}$0.5\% of constrained test-passing runs) barely move the aggregate statistics.

\textbf{Missed violations.}
If verifiers were to miss genuine constraint violations, the effect would be to preserve the A\% of non-compliant runs. This can only weaken the observed decay signal, making our reported constraint decay a conservative lower bound on the true effect. In other words, missed violations cannot inflate the decay, they can only mask it.

Together, these observations confirm that the constraint-decay findings are robust to verifier imperfections in both directions.

\begin{table}[h]
\centering
\footnotesize
\setlength{\tabcolsep}{3.6pt}
\begin{tabular}{@{}ll rr@{\;}c rr@{\;}c rr@{\;}c rr@{\;}c@{}}
\toprule
& & \multicolumn{3}{c}{\textbf{L0}} & \multicolumn{3}{c}{\textbf{L1}} & \multicolumn{3}{c}{\textbf{L2}} & \multicolumn{3}{c}{\textbf{L3}} \\
\cmidrule(lr){3-5} \cmidrule(lr){6-8} \cmidrule(lr){9-11} \cmidrule(lr){12-14}
\textbf{Agent} & \textbf{Model} & \textbf{A\%} & \textbf{Raw} & $\Delta$ & \textbf{A\%} & \textbf{Raw} & $\Delta$ & \textbf{A\%} & \textbf{Raw} & $\Delta$ & \textbf{A\%} & \textbf{Raw} & $\Delta$ \\
\midrule
Mini-SWE  & GPT-5-mini       & 51.7 & 51.7 & 0.0 & 46.8 & 46.8 & 0.0 & 27.1 & 27.7 & 0.6 & 23.7 & 25.1 & 1.4 \\
OpenHands & GPT-5-mini       & 65.8 & 65.8 & 0.0 & 63.2 & 63.2 & 0.0 & 56.7 & 57.7 & 1.0 & 52.2 & 54.7 & 2.5 \\
Mini-SWE  & Qwen3-Coder-Next      & 86.4 & 86.4 & 0.0 & 66.4 & 66.4 & 0.0 & 52.6 & 54.7 & 2.1 & 46.1 & 48.8 & 2.7 \\
OpenHands & Qwen3-Coder-Next      & 73.0 & 73.0 & 0.0 & 51.7 & 51.8 & 0.1 & 42.7 & 43.9 & 1.2 & 27.6 & 30.0 & 2.4 \\
Mini-SWE  & Qwen3-235B       & 29.6 & 29.6 & 0.0 & 10.7 & 10.7 & 0.0 &  9.1 &  9.1 & 0.0 &  2.3 &  2.9 & 0.7 \\
OpenHands & Qwen3-235B       & 26.2 & 26.2 & 0.0 & 17.7 & 17.7 & 0.0 &  3.1 &  3.1 & 0.0 &  0.8 &  0.8 & 0.0 \\
Mini-SWE  & Devstral-Small   &  6.8 &  6.8 & 0.0 &  4.7 &  4.7 & 0.0 &  2.3 &  3.6 & 1.3 &  2.9 &  3.0 & 0.1 \\
Mini-SWE & MiniMax-M2.5 & 90.5 & 90.5 & 0.0 & 79.3 & 79.3 & 0.0 & 75.0 & 77.0 & 1.9 & 66.8 & 68.9 & 2.0 \\
Mini-SWE & GPT-5.4 & 97.5 & 97.5 & 0.0 & 81.9 & 81.9 & 0.0 & 81.9 & 81.9 & 0.0 & 80.0 & 80.0 & 0.0 \\
\midrule
OpenHands & MiniMax-M2.5*    & 95.6 & 95.6 & 0.0 & 97.0 & 97.0 & 0.0 & 87.3 & 87.3 & 0.0 & 78.6 & 78.6 & 0.0 \\
Mini-SWE  & Kimi-K2.5*       & 85.4 & 85.4 & 0.0 & 70.9 & 72.5 & 1.6 & 62.9 & 62.9 & 0.0 & 53.7 & 53.7 & 0.0 \\
Mini-SWE  & GPT-5.2*         & 78.2 & 78.2 & 0.0 & 49.3 & 49.3 & 0.0 & 27.1 & 27.1 & 0.0 & 48.0 & 48.0 & 0.0 \\
\bottomrule
\end{tabular}
\caption{A\% with and without verifier enforcement. ``A\%'' zeros out assertion scores for runs that violate structural constraints; ``Raw A\%'' uses behavioral-test scores only. $\Delta$ = Raw A\% $-$ A\%. Full-set models use all 80 tasks; * = 16-task subset.}
\label{tab:verifier_comparison}
\end{table}

\section{Framework Leaderboard}
\label{app:framework_leaderboard}

Table~\ref{tab:framework_leaderboard} reports the full breakdown underlying Figure~\ref{fig:framework_sensitivity}: all eight frameworks across the eight model--scaffold configurations for which a full eight-framework sweep is available, with per-cell standard errors across individual runs.
Two configurations are omitted from the figure. Qwen3-235B is excluded because its near-zero scores (9.3\% and 10.3\% avg.\ A\%) leave all frameworks within noise, making its ranking uninformative.
The figure further shows the top-3 and bottom-3 frameworks by average A\%; the omitted aiohttp and Fastify rank fourth and fifth on every averaging basis (mid-tier configurations only, the six configurations in the figure, or all eight in the table below).

\begin{table}[h]
\centering
\footnotesize
\setlength{\tabcolsep}{3pt}
\begin{tabular}{@{}l cc cc cc c c c@{}}
\toprule
& \multicolumn{2}{c}{\textbf{GPT-5-mini}} & \multicolumn{2}{c}{\textbf{Qwen3-Coder-Next}} & \multicolumn{2}{c}{\textbf{Qwen3-235B}} & \textbf{MiniMax-M2.5} & \textbf{GPT-5.4} & \\
\cmidrule(lr){2-3} \cmidrule(lr){4-5} \cmidrule(lr){6-7} \cmidrule(lr){8-8} \cmidrule(lr){9-9}
& M & O & M & O & M & O & M & M & \textbf{Avg} \\
\midrule
Express & 60.0\,\tiny{$\pm$\,8.0} & 84.2\,\tiny{$\pm$\,5.9} & 71.8\,\tiny{$\pm$\,5.7} & 60.8\,\tiny{$\pm$\,7.6} & 14.9\,\tiny{$\pm$\,6.0} & 16.7\,\tiny{$\pm$\,5.3} & 73.6\,\tiny{$\pm$\,7.9} & 99.6\,\tiny{$\pm$\,0.2} & 60.2 \\
Koa & 49.2\,\tiny{$\pm$\,8.8} & 85.0\,\tiny{$\pm$\,6.4} & 66.4\,\tiny{$\pm$\,6.4} & 78.5\,\tiny{$\pm$\,4.7} & 11.3\,\tiny{$\pm$\,4.7} & 13.6\,\tiny{$\pm$\,4.2} & 81.4\,\tiny{$\pm$\,6.2} & 83.8\,\tiny{$\pm$\,6.3} & 58.7 \\
Flask & 37.8\,\tiny{$\pm$\,8.4} & 72.6\,\tiny{$\pm$\,7.3} & 74.9\,\tiny{$\pm$\,7.3} & 68.9\,\tiny{$\pm$\,7.3} & 29.4\,\tiny{$\pm$\,7.7} & 12.0\,\tiny{$\pm$\,4.8} & 76.6\,\tiny{$\pm$\,6.2} & 85.8\,\tiny{$\pm$\,6.3} & 57.2 \\
aiohttp & 34.7\,\tiny{$\pm$\,8.0} & 64.3\,\tiny{$\pm$\,8.1} & 48.0\,\tiny{$\pm$\,7.7} & 56.4\,\tiny{$\pm$\,7.2} & 11.1\,\tiny{$\pm$\,4.8} & 15.8\,\tiny{$\pm$\,6.0} & 72.4\,\tiny{$\pm$\,7.4} & 70.7\,\tiny{$\pm$\,8.2} & 46.7 \\
Fastify & 35.5\,\tiny{$\pm$\,7.2} & 55.8\,\tiny{$\pm$\,6.9} & 60.2\,\tiny{$\pm$\,4.8} & 28.2\,\tiny{$\pm$\,6.3} & 4.0\,\tiny{$\pm$\,2.3} & 6.5\,\tiny{$\pm$\,2.7} & 74.9\,\tiny{$\pm$\,5.0} & 66.1\,\tiny{$\pm$\,6.2} & 41.4 \\
Django & 33.7\,\tiny{$\pm$\,7.4} & 66.5\,\tiny{$\pm$\,8.1} & 32.1\,\tiny{$\pm$\,7.3} & 17.4\,\tiny{$\pm$\,5.4} & 0.7\,\tiny{$\pm$\,0.7} & 2.1\,\tiny{$\pm$\,1.4} & 68.9\,\tiny{$\pm$\,7.4} & 91.8\,\tiny{$\pm$\,4.6} & 39.1 \\
FastAPI & 21.4\,\tiny{$\pm$\,6.6} & 24.7\,\tiny{$\pm$\,7.2} & 63.1\,\tiny{$\pm$\,7.0} & 28.9\,\tiny{$\pm$\,7.0} & 3.7\,\tiny{$\pm$\,1.4} & 3.6\,\tiny{$\pm$\,1.7} & 77.0\,\tiny{$\pm$\,6.8} & 79.1\,\tiny{$\pm$\,8.3} & 37.7 \\
Hono & 6.1\,\tiny{$\pm$\,3.2} & 15.9\,\tiny{$\pm$\,6.4} & 54.0\,\tiny{$\pm$\,8.1} & 24.1\,\tiny{$\pm$\,7.0} & 6.9\,\tiny{$\pm$\,4.3} & 4.3\,\tiny{$\pm$\,2.6} & 85.1\,\tiny{$\pm$\,6.3} & 87.1\,\tiny{$\pm$\,5.7} & 35.4 \\
\midrule
\textbf{Spread} & 53.9 & 69.1 & 42.9 & 61.1 & 28.7 & 14.5 & 16.2 & 33.5 & \\
\bottomrule
\end{tabular}
\caption{Assertion pass rate A\% by framework, aggregated across constraint levels, with $\pm1$ standard error across individual runs. M\,=\,Mini-SWE-Agent; O\,=\,OpenHands. ``Spread'' is the best-to-worst difference over all eight frameworks; over the six frameworks shown in Figure~\ref{fig:framework_sensitivity} it is 53.9, 69.1, 42.9, 61.1, 28.7, 14.5, 16.2 and 20.5~pp respectively. All configurations cover the full 80-task set with 3 runs per task.}
\label{tab:framework_leaderboard}
\end{table}

\section{Generation Task Design}
\label{app:task_design}

Each generation task instructs an LLM agent to produce, from scratch, a fully functional REST API server that conforms to the RealWorld (Conduit) OpenAPI specification.  Tasks are assembled from a modular prompt template whose blocks are conditionally included depending on the target constraint level.

\subsection{Prompt Template}

The master template has the following structure (optional blocks are marked with~$\langle\cdot\rangle$):

\begin{quote}
\small\ttfamily
\{\,OpenAPI specification\,\}\\[4pt]
Generate a complete \{runtime\} REST API server compliant with the given specification.\\[4pt]
\#\# Requirements\\
-- You MUST use the \textbf{\{framework\}} framework.\\
-- You MUST work in the current directory.\\
$\langle$-- You MUST follow the architecture described below.$\rangle$\\
$\langle$-- You MUST use the database as described below.$\rangle$\\
$\langle$-- You MUST use the \{ORM\} ORM for handling the database.$\rangle$\\[4pt]
$\langle$\#\# Architecture\\
\textnormal{\textit{(Clean Architecture pattern: four layers, dependency rule, directory layout)}}$\rangle$\\[4pt]
$\langle$\#\# Database\\
\textnormal{\textit{(SQLite auto-schema \textbf{or} PostgreSQL with connection credentials)}}$\rangle$\\[4pt]
\#\# Mandatory Files\\
\textnormal{\textit{(requirements.txt/package.json + run.sh)}}\\[4pt]
\#\# Server Configuration\\
-- The server MUST listen on port \{port\}.\\
-- The server MUST expose GET /api/health-check returning 200.\\
-- All routes MUST be prefixed with /api.\\[4pt]
\#\# Evaluation Pipeline\\
\textnormal{\textit{(dependency install $\to$ run.sh $\to$ health-check poll $\to$ Postman test suite)}}
\end{quote}

The \texttt{\{runtime\}} placeholder is filled with \texttt{Python~3.12} or \texttt{Node.js}; \texttt{\{framework\}} with one of the eight target frameworks; \texttt{\{port\}} with \texttt{8080} for Python or \texttt{3000} for Node; and \texttt{\{ORM\}} with \texttt{SQLAlchemy} or \texttt{Sequelize}.  The OpenAPI specification at the top is a standard OpenAPI~3.0 YAML document (${\sim}$500 lines) defining all Conduit endpoints, schemas, and authentication requirements.

\subsection{Constraint Blocks}

The optional blocks correspond to the three constraint dimensions introduced in~\S\ref{sec:constraints}:

\textbf{Architecture.}  A paragraph prescribing the Clean Architecture pattern: four layers (routes/handlers, services/use cases, models/entities, repository/data access) with a strict top-down dependency rule (each layer may only import from the layer below) and a directory-per-layer layout.

\textbf{Database.}  Either a SQLite block (auto-create schema on startup) or a PostgreSQL block (auto-create schema on startup, with given fixed credentials: \texttt{user/password@postgres:5432/conduit}).

\textbf{ORM.}  A single sentence requiring SQLAlchemy (Python) or Sequelize (Node.js).

\noindent Constraint levels L0--L3 are formed by including an increasing number of these blocks:
L0 includes none (baseline);
L1 includes one (e.g., architecture only, or database only);
L2 includes two (e.g., architecture + database);
L3 includes all three.
For each of the 8 frameworks we generate all 10 valid constraint combinations, yielding 80 distinct generation tasks.

\subsection{Prompt Example}

Figure~\ref{fig:prompt_example} shows an abridged prompt for a task with all constraints active.

\begin{figure}[h]
\small
\begin{mdframed}[linewidth=0.5pt,innerleftmargin=6pt,innerrightmargin=6pt,innertopmargin=6pt,innerbottommargin=6pt]
\ttfamily
openapi: "3.0.1"\\
info:\\
\quad title: "Conduit API"\\
\quad version: "1.0.0"\\
paths:\\
\quad /api/users/login:\\
\quad\quad post: \textnormal{\textit{\ldots (19 endpoints, ${\sim}$500 lines total, omitted for brevity)}}\\[6pt]
\rmfamily
Generate a complete Python 3.12 REST API server compliant with the given specification.\\[4pt]
\textbf{\#\# Requirements}\\
-- You MUST use the \textbf{flask} framework.\\
-- You MUST work in the current directory.\\
-- You MUST follow the architecture described below.\\
-- You MUST use the database as described below.\\
-- You MUST use the SQLAlchemy Python ORM for handling the database.\\[4pt]
\textbf{\#\# Architecture}\\
You MUST follow the Clean Architecture pattern. Organize your code into these layers:\\
-- \textbf{Routes/Handlers layer}: HTTP request handling, input parsing, response formatting. No business logic.\\
-- \textbf{Services/Use Cases layer}: Business logic and orchestration. Framework-agnostic.\\
-- \textbf{Models/Entities layer}: Data structures and domain objects.\\
-- \textbf{Repository/Data Access layer}: All data storage operations. Accessed only through the Services layer.\\
Each layer must only depend on the layer below it. Keep each layer in its own directory.\\[4pt]
\textbf{\#\# Database}\\
Use \textbf{PostgreSQL} as the database. The database schema MUST be created automatically on server startup.\\
An instance of Postgres 16 is already running on port 5432.\\
Connection: username=user, password=password, db=conduit, host=postgres.\\[4pt]
\textbf{\#\# Mandatory Files}\\
1.~\textbf{requirements.txt} -- all pip dependencies.\\
2.~\textbf{run.sh} -- a shell script that starts the server.\\[4pt]
\textbf{\#\# Server Configuration}\\
-- The server MUST listen on \textbf{port 8080}.\\
-- The server MUST expose a health-check endpoint at \textbf{GET /api/health-check} returning HTTP 200.\\
-- All API routes MUST be prefixed with \textbf{/api}.\\[4pt]
\textbf{\#\# Evaluation Pipeline}\\
1.~\texttt{uv pip install --system -r requirements.txt}\\
2.~\texttt{chmod +x run.sh}\\
3.~\texttt{./run.sh} (starts your server in background)\\
4.~Health-check polls \texttt{GET http://localhost:8080/api/health-check} until 200.\\
5.~A Postman test suite runs against \texttt{http://localhost:8080/api}.
\end{mdframed}
\caption{Generation prompt for an L3 task.  The full OpenAPI specification (${\sim}$500 lines) is prepended verbatim but omitted here for space.}
\label{fig:prompt_example}
\end{figure}

\section{Feature Implementation Task Design}
\label{app:feature_task_design}

Feature implementation tasks evaluate an agent's ability to add missing functionality to an existing, partially stripped codebase.  Unlike generation tasks, the agent starts with a working repository and must read, understand, and extend it: simulating the common developer scenario of implementing a new feature in an established project.

\subsection{Construction by Codebase Ablation}

Each task is derived from a real open-source RealWorld (Conduit) implementation. Since the Conduit API is organized around well-defined feature groups (articles, comments, tags, profiles, users), we can cleanly remove an entire feature vertical (routes, controllers/handlers, services, data-access code, and model definitions) while leaving the rest of the codebase intact and functional. Concretely, for each repository we:

\begin{enumerate}[leftmargin=*,label=(\roman*)]
    \item Select one or more feature domains to ablate (e.g., ``comments,'' or ``articles + comments + tags'').
    \item Apply a \texttt{git diff} patch that removes all code belonging to those domains.
    \item Verify that the stripped repository still compiles, starts, and partially passes the test suite on the \emph{remaining} endpoints.
    \item Record the resulting task: the agent must restore full API compliance by re-implementing the removed features.
\end{enumerate}

Crucially, the number of files and lines affected by each task is not arbitrarily chosen: it is a direct consequence of how much code a given feature domain occupies across the repository's layered architecture.  A single-domain removal (e.g., comments) touches fewer files than a multi-domain removal (e.g., articles + comments + tags) because the latter spans more entities, more handlers, and more service logic.  This mirrors the real-world situation in which the scope of a feature implementation is determined by the feature itself.

\subsection{Task Inventory and Statistics}

We construct 20 tasks across five framework repositories (4 per framework), covering three ablation scopes summarized in Table~\ref{tab:feature_tasks_detail}.

\begin{table}[h]
\centering
\footnotesize
\begin{tabular}{@{}l l rr@{}}
\toprule
\textbf{Ablation scope} & \textbf{Tasks} & \textbf{Lines removed} & \textbf{Files} \\
\midrule
Single domain (comments)             & 5 & 176--639   & 4--15 \\
Single domain (profiles)             & 5 & 161--407   & 2--14 \\
Multi-domain (articles + tags $\pm$ comments) & 10 & 726--2{,}604 & 5--46 \\
\midrule
\textbf{All 20 tasks}                & 20 & 161--2{,}604 & 2--46 \\
\bottomrule
\end{tabular}
\caption{Feature implementation tasks grouped by ablation scope.  The wide range of lines and files reflects natural variation in how each repository organizes code across layers.}
\label{tab:feature_tasks_detail}
\end{table}

Table~\ref{tab:feature_repos} lists the five repositories and their provenance.

\begin{table}[h]
\centering
\footnotesize
\begin{tabular}{@{}ll c@{}}
\toprule
\textbf{Framework} & \textbf{Repository} & \textbf{Tasks} \\
\midrule
Express   & \texttt{realworld-express-prisma}      & 4 \\
Flask     & \texttt{realworld-flask}                & 4 \\
FastAPI   & \texttt{fastapi-realworld-backend}      & 4 \\
aiohttp   & \texttt{realworld-aiohttp}              & 4 \\
Hono.js   & \texttt{realworld-honojs}               & 4 \\
\bottomrule
\end{tabular}
\caption{Source repositories for feature implementation tasks.  All implement the full RealWorld Conduit API with PostgreSQL, an ORM (Prisma or SQLAlchemy), and a layered architecture.}
\label{tab:feature_repos}
\end{table}

\subsection{Prompt and Evaluation}

Each task is stored as a JSON file containing: (i)~the agent prompt (the full OpenAPI specification followed by a single instruction sentence), (ii)~a unified diff patch that strips the target feature before the agent starts, (iii)~repository metadata (URL, pinned commit, setup and run commands), and (iv)~task metadata (lines removed, files affected).

The evaluation pipeline clones the repository at the pinned commit, applies the ablation patch, runs the agent, then executes the same Postman test suite used for generation tasks. Because the ablation preserves all non-target features, the stripped repository already passes all tests for the retained endpoints; the agent's task is to bring the remaining (currently failing) tests back to passing.

Figure~\ref{fig:feature_prompt_example} shows the abridged prompt for an Express task.

\begin{figure}[h]
\small
\begin{mdframed}[linewidth=0.5pt,innerleftmargin=6pt,innerrightmargin=6pt,innertopmargin=6pt,innerbottommargin=6pt]
\ttfamily
openapi: "3.0.1"\\
info:\\
\quad title: "Conduit API"\\
\quad version: "1.0.0"\\
paths:\\
\quad /api/users/login:\\
\quad\quad post: \textnormal{\textit{\ldots (19 endpoints, ${\sim}$500 lines total, omitted for brevity)}}\\[6pt]
\rmfamily
Complete the existing repository and make it compliant to this OpenAPI specification by adding missing comments functionalities
\end{mdframed}
\caption{Example feature implementation prompt.  The agent receives the full OpenAPI specification followed by a single instruction.  The repository has been pre-stripped of 245 lines across 10 files (the entire comments vertical: route registrations, controller, Prisma schema entries, and validation middleware).}
\label{fig:feature_prompt_example}
\end{figure}

\section{Behavioral Test Suite}
\label{app:test_suite}

All task variants share a single specification compliant HTTP behavioral test suite, defined as a Postman\footnote{\url{https://www.postman.com/}} API testing framework collection and executed via the Newman\footnote{\url{https://github.com/postmanlabs/newman}} engine. The collection encodes a stateful integration-test scenario: requests run in a fixed order and later requests depend on state created by earlier ones (e.g., the authentication token, article slug, and comment identifier produced by prior steps).

\subsection{Coverage}

The suite contains 32~requests targeting 19~unique API endpoints (the full spec. surface) and 291~assertions.  Requests are organized into five sequential folders described in Table~\ref{tab:test_folders}.

\begin{table}[h]
\centering
\footnotesize
\begin{tabular}{@{}l r r l@{}}
\toprule
\textbf{Folder} & \textbf{Req.} & \textbf{Assert.} & \textbf{Scope} \\
\midrule
Auth                           &  5 &  30 & Register, login, token, profile update \\
Articles                       &  4 &  20 & Public listing (by author, tag, favorites) \\
Articles, Favorites, Comments  & 18 & 212 & Full CRUD + favorites + comments lifecycle \\
Profiles                       &  4 &  26 & Follow / unfollow social graph \\
Tags                           &  1 &   3 & Tag listing \\
\midrule
\textbf{Total}                 & 32 & 291 & \\
\bottomrule
\end{tabular}
\caption{Test suite folder breakdown.  The third folder accounts for 73\% of all assertions because it exercises the full article lifecycle including nested resources.}
\label{tab:test_folders}
\end{table}

\subsection{Assertion Taxonomy}

Each assertion belongs to one of four categories:

\textbf{Property presence}. Verifies that the JSON response contains expected keys (e.g., \texttt{user.email}, \texttt{article.slug}).

\textbf{Type validation}. Checks that the JSON response attribute values have the correct type (e.g., \texttt{articles} is an array, \texttt{favoritesCount} is an integer).

\textbf{Status-code checks}. Verifies the HTTP response code (e.g., \texttt{200~OK}, \texttt{201~Created}).

\textbf{State-transition checks}. Verifies that server-side state mutated correctly (e.g., \texttt{favorited} flips from \texttt{false} to \texttt{true} after the relative \texttt{POST} request).

\section{Token Consumption}
\label{app:token_consumption}

Table~\ref{tab:tokens_global} reports input and output token consumption aggregated across all runs of the full evaluation, broken down by constraint level. Table~\ref{tab:tokens_per_pair} provides the same breakdown per agent--model pair.

\begin{table}[h]
\centering
\small
\begin{tabular}{@{}l r r r r@{}}
\toprule
\textbf{Level} & \textbf{Avg Input} & \textbf{Avg Output} & \textbf{Total Input} & \textbf{Total Output} \\
\midrule
L0 & 787{,}553 & 11{,}334 & 185{,}074{,}973 & 2{,}663{,}400 \\
L1 & 1{,}558{,}208 & 19{,}940 & 958{,}298{,}049 & 12{,}263{,}053 \\
L2 & 1{,}983{,}938 & 21{,}768 & 1{,}599{,}053{,}697 & 17{,}545{,}404 \\
L3 & 2{,}271{,}671 & 23{,}696 & 983{,}633{,}532 & 10{,}260{,}232 \\
\midrule
Total & -- & -- & 3{,}726{,}060{,}251 & 42{,}732{,}089 \\
\bottomrule
\end{tabular}
\caption{Global token consumption by constraint level, aggregated across all agent--model configurations.}
\label{tab:tokens_global}
\end{table}

\begin{table}[h]
\centering
\footnotesize
\setlength{\tabcolsep}{4.7pt}
\begin{tabular}{@{}ll rrrr rrrr@{}}
\toprule
& & \multicolumn{4}{c}{\textbf{Avg Input Tokens}} & \multicolumn{4}{c}{\textbf{Avg Output Tokens}} \\
\cmidrule(lr){3-6} \cmidrule(lr){7-10}
\textbf{Agent} & \textbf{Model} & \textbf{L0} & \textbf{L1} & \textbf{L2} & \textbf{L3} & \textbf{L0} & \textbf{L1} & \textbf{L2} & \textbf{L3} \\
\midrule
Mini-SWE  & GPT-5-mini        &   19K &   50K &   69K &  124K &  4K &   9K &  10K &  12K \\
Mini-SWE  & GPT-5.2*          &   60K &  163K &  310K &  237K &  7K &  22K &  28K &  23K \\
Mini-SWE & GPT-5.4 & 74K & 79K & 101K & 125K & 5K & 6K & 8K & 9K \\
Mini-SWE & MiniMax-M2.5 & 364K & 845K & 1{,}009K & 1{,}254K & 10K & 19K & 20K & 22K \\
Mini-SWE  & Qwen3-Coder-Next       & 1{,}676K & 3{,}512K & 5{,}138K & 7{,}153K & 19K &  44K &  44K &  53K \\
Mini-SWE  & Qwen3-235B        &  107K &  167K &  324K &  415K &  6K &  10K &  12K &  13K \\
Mini-SWE  & Kimi-K2.5*        &  150K &  829K & 1{,}144K & 1{,}115K &  6K &  16K &  18K &  18K \\
\midrule
OpenHands & GPT-5-mini        &  304K &  519K &  908K & 1{,}732K & 21K &  24K &  27K &  35K \\
OpenHands & Qwen3-Coder-Next       & 3{,}774K & 6{,}560K & 7{,}310K & 7{,}115K & 23K &  31K &  33K &  32K \\
OpenHands & Qwen3-235B        &  702K & 1{,}267K & 1{,}373K & 1{,}474K & 11K &  18K &  20K &  20K \\
OpenHands & MiniMax-M2.5*     & 1{,}132K & 1{,}923K & 2{,}552K & 2{,}914K & 10K &  18K &  23K &  24K \\

\bottomrule
\end{tabular}
\caption{Average input and output tokens per run by constraint level and agent--model pair (K = $\times 10^3$, rounded). * = 16-task subset.}
\label{tab:tokens_per_pair}
\end{table}

\section{Execution Pipeline}
\label{app:pipeline}

This section details the end-to-end infrastructure that executes each task.  The pipeline is fully containerized and comprises two sequential phases: (i)~build (agent execution) and (ii)~evaluate (behavioral testing), each running in a pristine Docker environment to prevent state leakage between runs.

\subsection{Infrastructure Overview}

Each framework (e.g., Flask, Express) has a dedicated Docker Compose project defining three services on a shared network:

\textbf{conduit}. The application container.  Built from a multi-stage Dockerfile that (i)~installs two Python virtual environments for agent execution (one for Mini-SWE-Agent, one for OpenHands), and (ii)~clones the reference repository at a pinned commit with all framework dependencies pre-installed.  For generation tasks, the repository starts empty (only a \texttt{gitkeep} placeholder); for feature implementation tasks, it contains the full codebase.

\textbf{postgres}. A PostgreSQL~16 instance with fixed credentials (\texttt{user/password@postgres:5432/conduit}).

\textbf{newman}. A Node.js container with the Newman suite executor pre-installed, used exclusively during the evaluation phase.

\subsection{Build Phase (Agent Execution)}

The build phase proceeds as follows:

\begin{enumerate}[leftmargin=*,label=(\roman*)]
    \item \textbf{Image build.}  \texttt{docker compose build conduit} compiles the framework-specific image.
    \item \textbf{Container launch.}  The conduit container is started in detached mode with two bind-mounted volumes: \texttt{/agents} (agent scripts and model cost registry) and \texttt{/trajectories} (output directory for the agent's conversation trace).
    \item \textbf{Task preparation.}  The task JSON is copied into the container.  For feature implementation tasks, the ablation patch is applied via \texttt{git apply} to strip the target feature.  The Git history is then re-initialized and the baseline commit SHA is recorded.
    \item \textbf{Setup commands.}  Any task-specific setup commands run inside the container (e.g., \texttt{npm run migrate:develop} for Express/Prisma tasks, or \texttt{alembic upgrade head} for SQLAlchemy tasks).
    \item \textbf{Agent invocation.}  The agent script is executed with environment variables specifying the LLM model, API key, task prompt, and a system guidance message containing server startup instructions.  For example, Mini-SWE-Agent runs as:
    \begin{quote}\small\ttfamily
    /mini-env/bin/python /agents/mini\_swe\_sdk.py
    \end{quote}
    The agent can read, write, and execute code within the \texttt{/repository} folder.
    \item \textbf{Patch extraction.}  After the agent finishes, a \texttt{git diff} against the baseline SHA captures all changes as a unified diff patch, excluding generated artifacts (\texttt{node\_modules}, \texttt{\_\_pycache\_\_}, \texttt{*.db}, lock files).  This patch is the sole artifact carried to the evaluation phase.
    \item \textbf{Cleanup.} The container is stopped and removed; all volumes are destroyed.
\end{enumerate}

To avoid potential agentic loops, a maximum iteration limit is set for both agents: 300 turns for Mini-SWE-Agent and 200 turns for OpenHands.

\subsection{Evaluate Phase (Behavioral Testing)}

The evaluation phase applies the agent's patch to a \emph{pristine} container, ensuring that any side effects from the build phase (e.g., leftover processes, corrupted state) do not influence the result.

\begin{enumerate}[leftmargin=*,label=(\roman*)]
    \item \textbf{Pristine container.}  A new conduit container is started from the same image.  For feature implementation tasks, the ablation patch is applied first (reproducing the stripped baseline), followed by the agent's patch.
    \item \textbf{Setup and server start.}  Setup commands run again in the clean environment.  The server is started in the background via \texttt{run.sh}, with stdout/stderr redirected to a log file.
    \item \textbf{Health-check polling.}  The pipeline polls the health-check endpoint (\texttt{GET http://conduit:\{port\}/api/health-check}) up to 5~times at 5-second intervals.  If the server never responds, the evaluation continues to record a baseline measurement (typically 0\% assertions passed).
    \item \textbf{Newman test execution.}  The Postman collection is executed via Newman against the running server:
    \begin{quote}\small\ttfamily
    newman run Conduit.postman\_collection.json \textbackslash\\
    \quad --global-var APIURL=http://conduit:\{port\}/api \textbackslash\\
    \quad --reporters cli,html,csv
    \end{quote}
    \item \textbf{Artifact collection.}  Server logs, Newman reports (HTML + CSV), and agent patch are saved to the results directory.
    \item \textbf{Verifier execution.}  The three static verifiers (\S\ref{app:verifiers}) are run on the agent patch to assess structural compliance.
\end{enumerate}

\subsection{Walkthrough: Generation Task}

Consider the L3 task \texttt{flask-openapi-clean\_architecture-postgres-sqlalchemy}.

\begin{enumerate}[leftmargin=*,label=(\roman*)]
    \item The conduit container starts with an empty \texttt{/repository} directory (no pre-existing code).
    \item The agent receives the full prompt (Appendix~\ref{app:task_design}, Figure~\ref{fig:prompt_example}) and generates a Flask application from scratch: models, routes, services, repositories, configuration, \texttt{requirements.txt}, and \texttt{run.sh}.
    \item The build phase extracts a patch containing all created files (${\sim}$500--2000 added lines).
    \item The evaluate phase applies this patch to a fresh empty container, runs \texttt{uv pip install -r requirements.txt}, starts the server on port~8080, and executes the Postman suite.
    \item Verifiers check: (a)~$\geq$3 architectural layers with no upward imports, (b)~PostgreSQL evidence present and no SQLite evidence, (c)~SQLAlchemy evidence present.
\end{enumerate}

\subsection{Walkthrough: Feature Implementation Task}

Consider the Express task that ablates the comments feature (245 lines across 10 files).

\begin{enumerate}[leftmargin=*,label=(\roman*)]
    \item The conduit container starts with the full Express/Prisma repository at the pinned commit.  The ablation patch is applied, removing all comment-related code (routes, controller, Prisma schema fields, validation middleware).  After ablation, the server still starts and passes all non-comment tests.
    \item The agent receives the prompt (Appendix~\ref{app:feature_task_design}, Figure~\ref{fig:feature_prompt_example}) and must re-implement the missing comments functionality by reading the existing codebase and extending it.
    \item The build phase extracts a patch containing only the agent's additions.
    \item The evaluate phase starts from the pinned commit, applies the ablation patch, then applies the agent's patch on top, sets up the server execution (e.g., runs database migrations), starts the server on port~3000, and executes the Postman suite.
\end{enumerate}

In this case no verifiers are applied since the constraints are already implicitly stated in the codebase.

\section{Revised Agent Scaffolds}
\label{app:agent_scaffolds}

Both agent scaffolds required targeted modifications to operate in our greenfield generation setting and to ensure fully automated evaluation.

\paragraph{Mini-SWE-Agent.}
Mini-SWE-Agent is designed for issue-solving on existing repositories. To repurpose it for greenfield generation, we replaced the default ``fix the issue'' instance prompt with a generative workflow. The injected instructions direct the agent through four steps:
\begin{quote}
\small
\texttt{1. Analyze the codebase by finding and reading relevant files} \\
\texttt{2. Edit/Create the source code to solve your task} \\
\texttt{3. Test edge cases to ensure your solution is robust} \\
\texttt{4. Submit your changes and finish your work by issuing:} \\
\texttt{\ \ \ echo COMPLETE\_TASK\_AND\_SUBMIT\_FINAL\_OUTPUT}
\end{quote}
No other changes were made to the scaffold: the system prompt, action format, observation template, and execution environment remain identical to the upstream configuration.

\paragraph{OpenHands.}
OpenHands provides a built-in \texttt{finish} tool that the agent is expected to call when work is complete. In practice, however, several models prematurely halt execution by emitting a natural-language summary message instead of invoking the \texttt{finish} tool. This triggers the \texttt{FINISHED} status without the agent having completed the task, as OpenHands interprets the end of a \texttt{conversation.run()} cycle with a final agent message as a request for user input.

To mitigate this, we wrap the conversation loop with a re-prompting mechanism (up to 6 turns). After each \texttt{conversation.run()} call, we inspect the last event: if it is an agent message rather than a tool invocation, we send a follow-up prompt instructing the agent to either continue working or call the \texttt{finish} tool explicitly. The loop terminates only when the agent invokes \texttt{finish} or the maximum number of re-prompting turns is exhausted. This ensures that models prone to premature halting~\citep{ding2025nl2repo} are given the opportunity to resume work, yielding complete evaluation trajectories without manual intervention.

\end{document}